\documentclass[twocolumn]{revtex4}
\usepackage{graphicx}
\begin{document}

\title{Diversity of individual mobility patterns \\ and emergence of aggregated scaling laws}

\author{Xiao-Yong Yan$^{1,2,3}$}
\author{Xiao-Pu Han$^{4,5}$}
\author{Bing-Hong Wang$^{4}$}
\author{Tao Zhou$^{1,6\ast}$}

\affiliation{
$^{1}$Web Sciences Center, University of Electronic Science and Technology of China, Chengdu 611731, P.R. China\\
$^{2}$Department of Transportation Engineering, Shijiazhuang Tiedao University, Shijiazhuang 050043, P.R. China\\
$^{3}$Department of Systems Science, Beijing Normal University, Beijing 100875, P.R. China\\
$^{4}$Department of Modern Physics, University of Science and Technology of China, Hefei 230026, P.R. China\\
$^{5}$Institute of Information Economy and Alibaba Business College, Hangzhou Normal University, Hangzhou 310036, P.R. China\\
$^{6}$Microsoft Research Asia, Beijing 100080, P.R. China.\\
$^{\ast}$ Correspondence should be addressed to zhutou@ustc.edu
}

\begin{abstract}
 Uncovering human mobility patterns is of fundamental importance to the understanding of epidemic spreading, urban transportation and other socioeconomic dynamics embodying spatiality and human travel. According to the direct travel diaries of volunteers, we show the absence of scaling properties in the displacement distribution at the individual level,while the aggregated displacement distribution follows a power law with an exponential cutoff. Given the constraint on total travelling cost, this aggregated scaling law can be analytically predicted by the mixture nature of human travel under the principle of maximum entropy. A direct corollary of such theory is that the displacement distribution of a single mode of transportation should follow an exponential law, which also gets supportive evidences in known data. We thus conclude that the travelling cost shapes the displacement distribution at the aggregated level.
\end{abstract}

\maketitle

Positioning systems in mobile phones and vehicles and Wi-Fi devices in laptop computers and personal digital assistants have made quantitative analyses of human mobility patterns possible \cite{nat08,pnas12,pre09,acm06}. These analyses have a significant potential to reveal novel statistical regularities of human behavior, refine our understanding of the socioeconomic dynamics embodying spatiality and human mobility \cite{sci09,prp11}, and eventually contribute to controlling disease \cite{np11,prx11,pre09b,chao12}, designing transportation systems \cite{jtg01}, locating facilities \cite{pnas09}, providing location-based services \cite{Zheng2010,arx11a,acm11}, and so on.

Aggregated data from bank notes \cite{nat06}, mobile phones \cite{nat08} and onboard GPS measurements \cite{pre09} showed that the displacement distribution of human mobility, for both long-range travel and daily movements, approximately follows a power law. The scaling laws in long-range travel may result from the hierarchical organization of transportation systems \cite{pre11}, while the scaling laws in daily movements have recently been explained by the \emph{exploration and preferential return} mechanism \cite{np10}.

Thus far, we still lack solid results about human mobility patterns at the individual level. Inferring individual features from the aggregated data is very risky because the scaling law for the population could be a mixture of many individuals with different statistics \cite{pnas11}. In addition, the aforementioned data are not sufficient to draw conclusions at the individual level. First, data such as GPS records from taxis and the trajectories of bank notes consist of many individual movements, but these individuals are not easy to be distinguished from each other. Second, data such as GPS records from mobile phones and the trajectories of bank notes could not accurately capture purposeful travels with explicit origins and destinations. In fact, the displacement between two activations of a mobile phone may be just a tiny portion of a purposeful trip or a combination of several sequential trips, while the displacement between two registrations of a bank note could be the result of a number of sequential trips made by different people.

Instead of using proxy data, we analyze the travel diaries of hundreds of volunteers. Though the data set is small, it contains personal profiles and explicit positions of origins and destinations, allowing quantitative and authentic analyses at the individual level. In contrast to the scaling laws in aggregated data, individuals show diverse mobility patterns, and few of them display the scaling property. In fact, the trajectories of students and employees are dominated by trips connecting homes with schools and workplaces, respectively, while trips are distributed more homogeneously among different locations for others such as retirees, homemakers and unemployed people. The aggregated displacement distribution follows a power law with an exponential cutoff, which can be analytically explained by the mixed nature of human travel under the principle of maximum entropy. In addition, this theory predicts that the displacements using a single mode of transportation will follow an exponential distribution, which is also supported by the empirical data on taxi trips, car trips, bus trips and air flights.

\section*{\large Results}

\subsection*{Individual mobility patterns.}

Our analysis of human mobility is based on a data set of 230 volunteers' six-week travel diaries in Frauenfeld, Switzerland \cite{jts05}. This data set contains the volunteers' personal information, including age, job and sex, and 36761 trip records. By calculating the spherical distance between the origin and destination from their longitudes and latitudes, we can obtain the length of each trip (see details about data in \textbf{Methods}).

We first measure the individual displacement distributions from the data set. Figs. 1(a)-1(c) show three typical individuals' displacement distributions (Table S1 presents all volunteers' displacement distributions), from which we cannot find any universal scaling properties. Indeed, when we use the {\it Kolmogorov-Smirnov} test \cite{sam09} to test whether the distributions fit power laws, we find that 87.8\% of the individuals cannot pass the test (statistical validation results are listed in Table S2, and the details about {\it Kolmogorov-Smirnov} test are shown in \textbf{Methods}). This result strongly suggests the absence of scaling laws in human travel at the individual level.

To reveal the underlying structure of individual trips, we assign to each individual a mobility network, in which nodes denote locations visited by individuals, edges represent the trips between nodes and edge weight is defined as the number of corresponding trips \cite{sci10}. Figs. 1(d)-1(f) show three typical individuals' mobility networks (all networks are presented in Table S1). As shown in Fig. 1 and Table S1, for most students and employees, their edge weights are highly heterogeneous. For each individual, we call the trip corresponding to the edge with the largest weight the {\it dominant trip} and  define the domination ratio $d$ as the ratio of the weight of the dominant trip to the total weight. Fig. 2 reports the distribution of domination ratios for different groups of individuals, from which we can see that the student group has the largest $d$ on average and the employees' average domination ratio is smaller than that of the students but larger than that of the other group.

The difference of $d$ results from the fact that students and employees frequently travel between homes and schools/workplaces in working days but retirees or homemakers do not have to do so. The peak values in the displacement distributions of students and employees are thus usually determined by the lengths of their dominant trips. Because the lengths of dominant trips are not necessarily small, the displacement distribution for an individual is usually not right-skewed and is far different from a power law. In addition, the significant role of the dominant trip indicates that an individual's traveling process in general cannot be characterized by the L\'evy flight \cite{nat06} or truncated L\'evy flight \cite{nat08}.

\begin{figure}
\includegraphics[width=8.6cm]{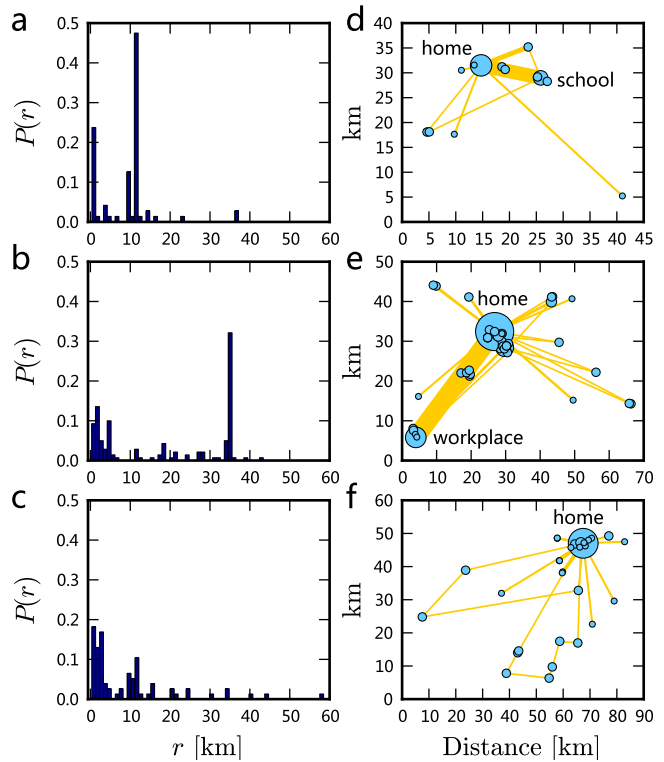}
    \caption{\label{fig1}{\bf Individual mobility patterns.}  (a-c) Displacement distributions for three typical individuals ((a) a student, (b) an employee, (c) a retiree), where the peak values for the student and the employee result from the trips between two most frequently visited locations. (d-f) Mobility networks for the three individuals, where the area of a node is proportional to its number of visits and the width of an edge is proportional to its weight.}
\end{figure}

\begin{figure}
\includegraphics[width=8.6cm]{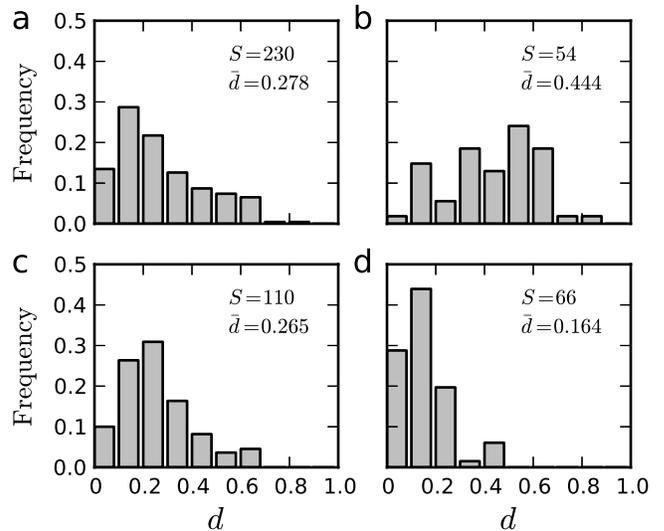}
    \caption{\label{fig2}{\bf Distribution of the domination ratios.}  (a) Population.  (b) Student group.  (c)  Employee group. (d) Others. $S$ is the number of group members, and $\bar{d}$ is the average domination ratio.}
\end{figure}

\begin{figure}
\includegraphics[width=8.6cm]{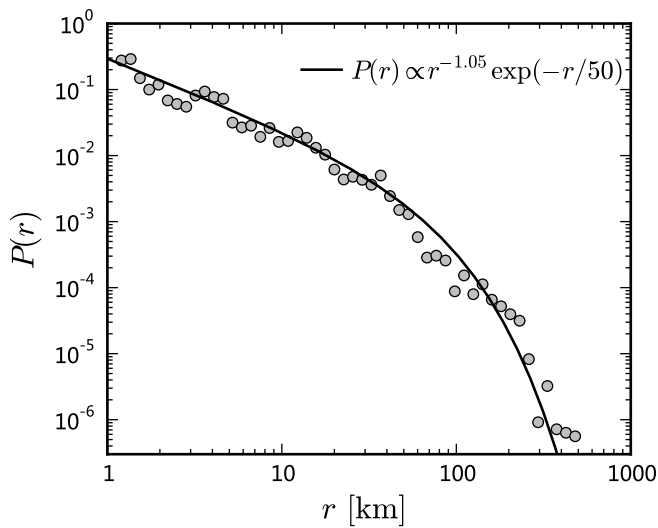}
    \caption{\label{fig3}{\bf Displacement distribution $P(r)$ of the aggregated data.} The solid line indicates a power law with an exponential cutoff. The data were binned using the logarithmic binning method (see \textbf{Methods} for details).}
\end{figure}

\begin{figure}
\includegraphics[width=8.6cm]{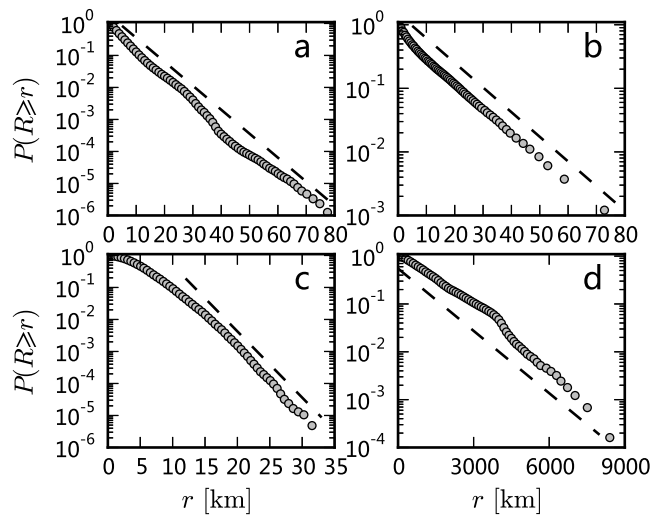}
    \caption{\label{fig4}{\bf Cumulative displacement distributions for a single mode of transportation.} (a) 12,028,929 taxi passenger trajectories in Beijing.  (b) 46,541 car trips in Detroit.  (c) 783,210 bus trips in Shijiazhuang.  (d) 205,534 air-flight travels in US.}
\end{figure}

\subsection*{Scaling property in aggregated data.}
The aggregated displacement distribution of individuals (see Fig. 3) is well approximated by a power law with an exponential cutoff $P(r)\propto  r^{-1.05}\exp({-{r / 50}})$ (the fitness significance $p$-value by the {\it Kolmogorov-Smirnov} test \cite{sam09} is 1.000 and the standard {\it Kolmogorov-Smirnov} distance $D$ is 0.039, see \textbf{Methods} and Fig. S1 for details), which is similar to those observed for bank notes \cite{nat06} and mobile phone users \cite{nat08}. As shown above, this scaling property is not a simple combination of many analogous individuals. We assume that the total travel cost is $C$, the number of trips with cost $c_i$ is $n_i$. According to the \emph{maximum entropy principle} \cite{sm75}, the two constraints, $\sum{n_{i}}=N$ and $\sum{n_{i} c_{i}}=C$, lead to the solution $n_i\propto \exp(-c_i/\bar{c})$, where $\bar{c}=C/N$ is the average travel cost. Denote the density of trips with cost $c$ by $P(c)$, then $P(c)\propto \exp(-c/\bar{c})$.

The travel cost is commonly approximated as the weighted sum $c\approx \eta t+\mu m$, where $\eta$ and $\mu$ are two coefficients, and $t$ and $m$ are the costs involving time and money, respectively. Previous empirical studies have suggested that the monetary cost is approximately proportional to the travel distance as $m\approx \nu r$ \cite{htm00}, while the travel time approximately obeys a hybrid form $t\approx \phi \ln r + \omega r + \psi$ \cite{ars99,geo05}, where $\nu$, $\phi$, $\omega$ and $\psi$ are coefficients. The logarithmic term results from the mixture of modes of transportation \cite{lac05}. Apparently, people move faster when traveling longer distances: we walk from classroom to office but take an airplane from US to China. Figure S2 reports the statistics related to travel times of the data set used in this paper. Although the data set is not large enough and contains some noisy points, overall speaking, the travel time $t$ grows in a hybrid form as mentioned above, with $\phi \approx 9$ and $\omega \approx 0.4$.

Integrating the aforementioned terms, we obtain the displacement distribution $P(r)\propto  (\beta / r + 1/\kappa)r^{-\beta}\exp(-r/\kappa)$, where $\beta = \eta\phi / \bar{c}$ and $\kappa = \bar{c} / (\mu \nu + \eta \omega)$. When $\kappa$ is large, the distribution is approximated as a power-law with an exponential cutoff. Indeed, for the real data, $\kappa=40$ and $\beta=0.38$, so $\frac{\beta}{r}\gg \frac{1}{\kappa}$ for $r<100$, that is, the term $\frac{1}{\kappa}$ can be neglected. As shown in Fig. S3, the corresponding fitting line is very close to a power law with an exponential cutoff (but with a slightly higher power-law exponent 1.38).

A direct corollary of maximum entropy principle is that the displacement distribution should follow an exponential form if it only accounts for trips from a single mode of transportation because in that case, $c \propto r$. This corollary gets supportive evidences from a number of empirical studies on disparate systems \cite{Bazzani2010,pls11,arx11,psa12,ijmpc,pls12} (Bazzani \emph{et al.} \cite{Bazzani2010} observed a slight deviation from the exponential law). Fig. 4 reports empirical cumulative distributions for taxi trajectories in Beijing \cite{psa12}, car trips in Detroit (downloaded from \emph{www.semcog.org}), bus trips in Shijiazhuang (collected by the authors) and air flights in the US \cite{arx11}. The probability density distributions are shown in Fig. S4. All distributions can be well characterized by exponential-like functions.

\section*{\large Discussion}

\noindent The general lessons that we learned from the present analysis could be used to refine our knowledge of human mobility patterns. The displacement distributions for aggregated data usually display power-law decay with an exponential cutoff. Meanwhile, there are examples ranging from taxi trips to air flights in which the displacement distributions are exponential. In these examples, every displacement distribution is generated  by trips involving a single mode of transportation, which corresponds to a linear relation between the travel cost and distance and eventually results in an exponential displacement distribution according to the principle of maximum entropy. In a word, we believe the travel cost is one main reason resulting in the regularities in aggregated statistics. The present results suggest that the form (power law or exponential or other) of deterrence function in the gravity law for human travel \cite{nat12} may be sensitive to the modes of transportation under consideration.

This study warns researchers of the risk of inferring individual behavioral patterns directly from aggregated statistics. Analogously, the temporal burstiness of human activities is widely observed, and the researchers are aware of the fact that the aggregated scaling laws could either be a combination of a number of individuals, each of whom displays scaling laws similar to the population \cite{nat05}, or the result of a mixture of diverse individuals, most of whom exhibit far different statistical patterns than the population \cite{pnas08,psa06,pnas10}. In comparison, such issues are less investigated for spatial burstiness. In particular, experimental analyses on individuals has rarely been reported. Determining whether the displacement distribution of an individual follows a power-law distribution will require further data and analysis.

It is already known to the scientific community that a number of Poissonian agents with different acting rates can make up a power-law inter-event time distribution at the aggregated level \cite{pnas08,psa06,pnas10}, and very recently, Proekt \emph{et al.} \cite{Proekt2012} showed that the aggregated scaling laws on inter-event time distribution may be resulted from different time scales. Petrovskill \emph{et al.} \cite{pnas11} have applied similar (yet different) idea in explaining the aggregated scaling laws in walking behavior. Although being mathematically and technically different, this work embodies some similar perspectives, because the different transportation modes indeed assign different scales onto space: the world becomes smaller by air flights while a city is really big by walking. Elegant analogy between temporal and spatial human behaviors will benefit the studies of each other.

Many known mechanisms underlie the scaling laws of complex systems \cite{im04,cp05,prp11b}, including rich get richer \cite{bio55,sci99,Lu2013}, good get richer \cite{np07,Zhou2011}, merging and regeneration \cite{epjb05}, optimization \cite{epl02,eco05},  Hamiltonian dynamics \cite{pre03}, stability constraints \cite{prl09}, and so on. The individual mobility model by Song {\it et al.} \cite{np10} is a typical example embodying the rich get richer mechanism. We have implemented such model. As shown in Fig. S5, the exploration and preferential return model can well reproduce the diversity of individual mobility patterns. In addition, for this model, the Gibbs entropy of the displacement distribution at the individual level increases continuously due to the increasing number of locations as well as links connecting location pairs. However, the exploration and preferential return model does not explain why the lengths of exploration trips should follow a power law, which is a core assumption leading to the power-law-like aggregated displacement distribution. Therefore, our work has complemented Ref. \cite{np10} and other related works in two aspects: (i) providing supportive empirical observation at the individual level; (ii) providing alternative explanation on the emergence of scaling in aggregated displacement distribution. Very recently, from the analysis on mobility patterns in an online game, Szell {\it et al.} \cite{sp12} observed a characteristic jump length and guessed that the existence of the characteristic length may be due to the single mode of transportation. The present theory could explain their observation since a jump in such online game costs time that is proportional to the jump length.

\section*{\large Methods}

\noindent  \textbf{Data description.} This work was performed using a travel survey data set that contains 230 volunteers' six-week travel diaries in Frauenfeld, Switzerland \cite{jts05}. The survey was conducted among 230 volunteers from 99 households in Frauenfeld and the surrounding areas in Canton Thurgau from August to December 2003. The volunteers reported their daily travel by filling out (paper and pencil based) self-administrated questionnaire day by day in a six-week period. Each reported trip includes the information of origin, destination and purpose. The origin and destination of a trip were geocoded by longitude and latitude. The quality of the geocoding is very high - with 60\% of trips captured within 100 m of their true origins and destinations and 90\% within 500 m. The purpose of trip was classified into work, shopping, education, home, leisure, business and other. The data has been cross-checked to ensure the consistency and filtered to remove outliers as well as unclear and omitted destination addresses. The final cleaned data set includes 36761 trip records. Besides, the data set also contains socio-demographic characteristics of the volunteers' personal information such as age, job and sex.

\noindent \textbf{Kolmogorov-Smirnov (KS) test.}  Given an observed distribution $P(x)$, we firstly assume that it obeys a certain form $F(x;a_1,a_2,\cdots,a_l)$, with a set of parameters $a_1,a_2,\cdots,a_l$, whose values are estimated by using the maximum likelihood method \cite{sam09}. The standard KS distance is defined as the maximal distance between the cumulative density functions of the observed data $P^c(x)$ and the fitting curve $F^c(x)$, namely $D^{KS}_{real}=\max_x\left| P^c(x)-F^c(x) \right|$. We independently sample a set of data points according to $F^c(x)$, such that the number of sampled data points is the same as the number of observed data points, and then calculate the maximal distance (denoted by $D^{KS}_{sample}$) between $F^c(x)$ and the cumulative density function of the sampled data points. The $p$-value is defined as the probability that $D^{KS}_{real}<D^{KS}_{sample}$. In this paper, we always implement 1000 independent runs to estimate the $p$-value.

\noindent  \textbf{Logarithmic binning.} The statistical nature of sampling will lead to the increasing noise in the tails of empirical power-law-type distributions. Applying the procedure of logarithmic binning \cite{logbin} can smooth the noisy tail. Logarithmic binning is a procedure of averaging the data that fall in the specific bins whose size increases exponentially. For each bin the observed value are normalized by dividing by the bin width and the total number of observations (see Fig. 3).

\section*{ACKNOWLEDGMENTS} 
This work was partially supported by the National Natural Science Foundation of China (NNSFC) under Grant Nos. 11222543, 11205040, 11275186 and 91024026, Program for New Century Excellent Talents in University under Grant No. NCET-11-0070, and Huawei University-Enterprise Cooperation Project under Grant No. YBCB2011057.

\end{document}